\documentclass[aps,prl,twocolumn,amsmath,amssymb]{revtex4}

\usepackage{graphicx}
\usepackage{dcolumn}
\usepackage{bm}
\usepackage{lineno}

\newcommand{\hp}{(C$_5$H$_{12}$N)$_2$CuBr$_4$}

\begin{document}


\preprint{APS/123-QED}

\title{Diverging thermal expansion of the spin-ladder system
(C$_5$H$_{12}$N)$_2$CuBr$_4$}

\author{T.\ Lorenz$^1$, O.\ Heyer$^1$, M.\ Garst$^2$, F.\ Anfuso$^2$, A. Rosch$^2$,
 Ch.\ R\"{u}egg$^3$, and K. Kr\"{a}mer$^4$}

\affiliation{ $^1$II.\,Physikalisches Institut, Universit\"{a}t zu
K\"{o}ln, Z\"{u}lpicher Stra{\ss}e 77, 50937 K\"{o}ln, Germany \\
$^2$Institut f\"{u}r Theoretische Physik, Universit\"{a}t zu
K\"{o}ln, Z\"{u}lpicher Stra{\ss}e 77, 50937 K\"{o}ln, Germany \\
$^3$London Centre for Nanotechnology and Dep. of Phys. and
Astronomy,
University College London WC1E 6BT, UK\\
$^4$Department of Chemistry and Biochemistry, University of Bern,
Freiestrasse, CH--3000 Bern 9, Switzerland}

\email{lorenz@ph2.uni-koeln.de}

\date{\today}

\begin{abstract}
We present high-resolution measurements of the $c^\star$-axis
thermal expansion and magnetostriction of piperidinium copper
bromide \hp. The experimental data at low temperatures is well
accounted for by a two-leg spin-ladder Hamiltonian. The thermal
expansion shows a complex behavior with various sign changes and
approaches a $1/\sqrt{T}$ divergence at the critical fields. All
low-temperature features are semi-quantitatively explained within
a free fermion model; full quantitative agreement is obtained
with Quantum Monte Carlo simulations.
\end{abstract}


\maketitle

In recent years there has been an increasing interest in spin
compounds exhibiting magnetic field-induced quantum phase
transitions. Examples are a Bose-Einstein condensation of magnons
\cite{Affleck90,Giamarchi99} studied in three-dimensional (3D)
systems like coupled spin dimers
\cite{Ruegg03,Johannsen05,Lorenz07} or arrays of coupled spin-1
chains \cite{Zapf1, Zapf2} or spin-ladders
\cite{watson01a,Garlea}. These systems share a zero-field ground
state with a spin gap. In a magnetic field two quantum phase
transitions are present; at $H_{c1}$ the gap closes and at
$H_{c2}$ the fully field-polarized state is reached. An
interesting scenario arises in the limit of very weakly coupled
ladders or chains where an extended temperature regime controlled
by 1D physics is expected. The low-dimensionality makes the
investigation of quantum critical properties particularly
exciting, and the thermal expansion $\alpha$ is especially suited
for this purpose. In fact, it has been shown that the
Gr\"{u}neisen parameter $\Gamma = \alpha/C$, where $C$ is the
specific heat, necessarily diverges at a quantum phase transition
\cite{Zhu03}. In addition, $\alpha$ shows a sign change whose
location in the phase diagram indicates accumulation of entropy
\cite{Garst05}. For $D<1/\nu$, where $\nu$ is the correlation
length exponent, even $\alpha$ will diverge at criticality. For a
spin-ladder, $\nu = 1/2$, the thermal expansion at the critical
fields is expected to behave as $\alpha \sim 1/\sqrt{T}$
\cite{Zhu03}. Experimentally, this is largely unexplored due to
the lack of suitable model materials.

In this letter we present a study of the thermal expansion and
the magnetostriction of single crystalline piperidinium copper
bromide \hp\ whose magnetic subsystem is a very good realization
of a two-leg spin ladder~\cite{watson01a} with Hamiltonian
\begin{align}\label{modelhamiltonian}
\mathcal{H} &= \sum_{i=1}^N \left[ J_\perp \mathbf{S}_{i,1}
\mathbf{S}_{i,2} + J_\parallel \left(\mathbf{S}_{i,1}
\mathbf{S}_{i+1,1} + \mathbf{S}_{i,2} \mathbf{S}_{i+1,2} \right)
\right. \nonumber\\ &\hspace{3em} \left. - g \mu_B H
\left(S^z_{i,1} + S^z_{i,2} \right)\right] \, .
\end{align}
The critical fields $H_{c1} = 6.8$~T and  $H_{c2} = 13.9$~T imply
for the exchange couplings the values $J_\perp/k_B = 12.9$ K  and
$J_\parallel/k_B = 3.6$ K \cite{HcFormulae}. The inter-ladder
couplings are very weak since no indications of three-dimensional
magnetic ordering are present down to $T\simeq
100$~mK~\cite{ruegg}. On approaching each of the critical fields,
we find highly anomalous temperature dependencies of $\alpha$.
For $0.3~{\rm K}\lesssim T \lesssim 2$~K, $\alpha(T)$ approaches
$1/\sqrt{T}$ divergences with opposite signs for $H_{c1}$ and
$H_{c2}$ in agreement with the expected quantum critical
behavior. Away from the critical fields, $\alpha$ shows a rather
complex structure with various sign changes, essentially
antisymmetric with respect to $(H_{c1}+H_{c2})/2$. All these
low-temperature features are reproduced semi-quantitatively
within a model of free fermions, although the latter is known to
be valid only in the vicinity of $H_{c1/c2}$.

\hp\ crystallizes in a monoclinic structure~\cite{patyal90a}. The
legs of the spin ladders are oriented along the $a$ axis and the
rungs roughly ($\approx 20^\circ$) along $c^\star$. The single
crystals used in this study have been grown from
solution~\cite{ruegg}. We present high-resolution measurements of
the thermal expansion $\alpha(T)=1/L_0 \cdot \partial \Delta
L(T)/\partial T $, magnetostriction $\varepsilon(H)=\Delta
L(H)/L_0$ and its field derivative $\lambda = \partial
\varepsilon(H)/\partial H$. Here, $L_0$ is the sample length
along $c^\star$; $\Delta L(T)$ and $\Delta L(H)$ denote the
temperature- and field-induced change at constant $H$ and $T$,
respectively.  The measurements have been performed on a
home-built capacitance dilatometer in magnetic fields $H\|
c^\star$ up to 17~T for $0.3~{\rm K} \lesssim T \lesssim 10$~K.

\begin{figure}
\includegraphics[width= \linewidth]{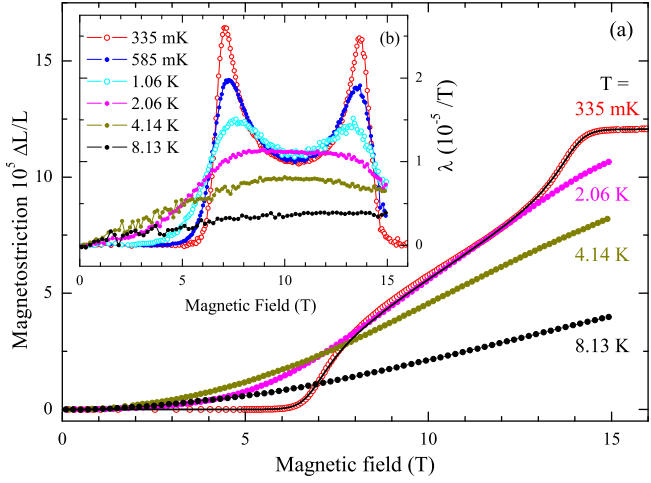}
\caption{\label{MS} (a) Magnetostriction $\varepsilon=\Delta
L(H)/L$ and (b) $\lambda=\partial \varepsilon/\partial H$
measured along $c^\star$ at various temperatures. The solid line
in (a) is a fit of $\varepsilon$ at 335~mK within a free fermion
model.}
\end{figure}

Fig.~\ref{MS} displays $\varepsilon(H)$ measured at various
constant temperatures. At $T=335$~mK, $\varepsilon(H)$ is field
independent up to about 6~T, then it continuously increases until
it saturates for $H \gtrsim 15$~T. This behavior strongly
resembles that of the low-temperature
magnetization~\cite{watson01a}. The inset of Fig.~\ref{MS} shows
the magnetostriction coefficient $\lambda(H)$ with two pronounced
peaks at 335~mK, whose magnitude shrink with increasing
temperature until around 2~K both peaks merge into one broad
plateau which further broadens towards higher $T$. In order to
identify the critical fields specified above we used linear
interpolations of the peak positions of $\lambda(H)$ towards
$T=0$~K.

\begin{figure}
\includegraphics[width= .9\linewidth]{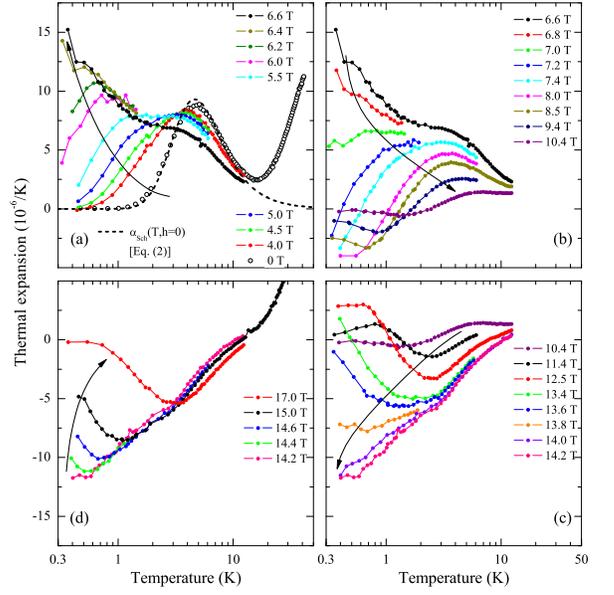}
\caption{\label{tad} {Thermal expansion $\alpha=1/L \cdot \partial
L/\partial T$ measured along the $c^\star$ direction in various
magnetic fields (a) below $H_{c1}$, (b,c) between $H_{c1}$ and
$H_{c2}$, and (d) above $H_{c2}$. Note the logarithmic
temperature scale. The dashed line in (a) is a fit with the
Schottky formula Eq.~(\ref{schottky}). The arrows signal the
order of the $\alpha(T)$ curves with increasing field.}}
\end{figure}

In Fig.~\ref{tad}(a) we summarize the thermal expansion data for
$H\lesssim H_{c1}$. In zero field, $\alpha(T)$ has a pronounced
peak around 5~K followed by a strong increase above about 15~K.
The latter is due to the usual thermal expansion of phononic
origin, while the 5~K peak is well described by a Schottky
formula indicating the thermal occupation of triplet excitations,
see discussion below. These degenerate triplets are split upon
increasing magnetic field which is reflected in a broadening and
slight shift of the Schottky peak towards lower $T$. For $H
\gtrsim 5.5$~T, the magnitude of the peak starts to grow until
around $H_{c1}$ a continuous increase of $\alpha(T)$ down to our
lowest $T\simeq 0.3$~K is observed. For $H>H_{c1}$, this low-$T$
increase rapidly weakens and $\alpha$ becomes even negative for
$H=7.2$~T, see Fig.~\ref{tad}(b). With further increasing field
this sign change shifts towards higher $T$ and a clear
minimum-maximum structure of $\alpha(T)$ becomes visible, whose
amplitude continuously decreases until a comparatively
featureless $\alpha(T)$ curve is reached at the aforementioned
symmetry field $(H_{c1}+H_{c2})/2=10.4$~T. On approaching
$H_{c2}$ we observe the same systematics of $\alpha(T)$ but with
inverted signs, Fig.~\ref{tad}(c); in particular, a negative
Schottky peak develops above $H_{c2}$, see Fig.~\ref{tad}(d).

\begin{figure}
\includegraphics[width= 0.7\linewidth]{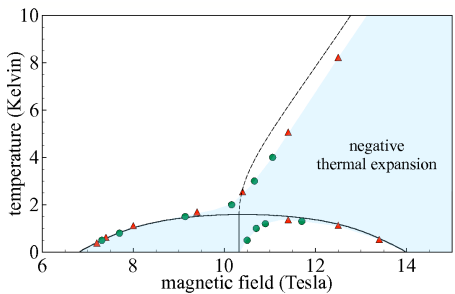} 
\caption{\label{zeros} Regions of positive and negative thermal
expansion $\alpha$ along the $c^\star$ axis of \hp . The
triangles and circles denote the positions of vanishing $\alpha$
determined from an interpolation along the $T$ and $H$ axis,
respectively. In the shaded area $\alpha$ is negative. The dashed
line indicates the zeros of $\alpha$ as predicted by the Schottky
formula (\ref{schottky}). The solid lines show zeros of $\alpha$
derived from the free-fermion fitting formula (\ref{alphaFF}),
which for $T=0$ start at the two quantum critical points located
at $H_{c1} = 6.8$~T and $H_{c2} = 13.9$~T and at
$(H_{c1}+H_{c2})/2$. Note that a crossing point of lines of zeros
as suggested by Eq.~(\ref{alphaFF}) is not realized in the
experiment as the regions of negative thermal expansion are
connected.}
\end{figure}

We now turn to the theoretical discussion of the magnetic
contributions to the thermal expansion and magnetostriction
arising from the Hamiltonian (\ref{modelhamiltonian}). As
$J_\parallel/J_\perp \approx 1/4 $ for \hp\ we can resort to the
strong coupling limit $J_\parallel \ll J_\perp$ where the physics
emerges in a transparent way. In this limit, the Hamiltonian
(\ref{modelhamiltonian}) represents a sum of weakly interacting
dimers in 1D. For $k_B T \gg J_\parallel$, the magnetic
contribution to $\alpha$ is then attributed to the thermal
occupation of single-dimer states described by the Schottky
formula
\begin{equation}
\alpha_{Sch} =
\frac{\gamma e^{\beta(h+J_\perp)} \left[ h \left( 1 - e^{2\beta h} \right)
 + J_\perp \left(1 + e^{\beta h} +
 e^{2 \beta h}\right)\right]}{k_B T^2 \left(1 +e^{\beta h}+ e^{2 \beta h}
 + e^{\beta(h+J_\perp)}\right)^2}
\label{schottky}
\end{equation}
where $1/\beta = k_B T$ and $\gamma = \frac{1}{V_D}
\frac{\partial J_\perp}{\partial p}$ with the volume per dimer
$V_D \simeq 859$~\AA$^3$~\cite{ruegg}. A fit for $H=0$ (dashed
line in Fig.~\ref{tad}(a)) yields $\gamma \simeq 11.7\times
10^{-5}$ \cite{HcFormulae2} implying a uniaxial pressure,
$p_{c^\star}\| c^\star $, dependence $\partial \ln
J_\perp/\partial p_{c^\star} \simeq 55$\%/GPa. In lowest order,
the intra-chain coupling just leads to a mean-field shift of the
effective magnetic field, $h = g \mu_B H - J_\parallel M/(g
\mu_B)$, where $M$ is the magnetization of a single dimer. The
Schottky formula (\ref{schottky}) accounts well for the field
dependence of the 5~K peak (not shown). In particular, it
identifies for any given temperature a unique magnetic field
where the thermal expansion vanishes and changes sign, see
Fig.~\ref{zeros}. Upon decreasing temperature, the positions of
vanishing $\alpha$ in the $(H,T)$ phase diagram shift to lower
fields towards the location of the singlet-triplet transition, $h
\simeq J_\perp$.

However, when temperature is of order $k_B T \simeq J_\parallel$
the behavior changes qualitatively and the Schottky formula
ceases to be valid. In contrast to a simple singlet-triplet level
crossing of isolated dimers, the kinetic energy of the triplet
excitations gives rise to an extended gapless phase at zero
temperature terminated by two quantum critical points at
$H_{c1/c2}$. In the low-$T$ limit $k_B T \ll J_\perp$, the
Hamiltonian (\ref{modelhamiltonian}) can be mapped onto an
effective XXZ spin-1/2
chain~\cite{Chaboussant98,Mila98,Totsuka98} that is described by
interacting tight-binding Jordan-Wigner fermions with bandwidth
$J_\parallel$ and chemical potential tuned by magnetic field,
$\mu = g \mu_B H - J_\perp  + \mathcal{O}(J_\parallel)$. Here,
the magnetic field and the pressure dependence $J_\perp(p)$ only
enter via the chemical potential $\mu$ which implies essentially
the same behavior for the derivatives of the free energy with
respect to $H$ and $p$. This not only explains the close
correspondence between magnetization \cite{watson01a} and
$\varepsilon(H)$ of Fig.~\ref{MS} in the low-$T$ limit but it
also provides an intuitive interpretation of the various sign
changes of the thermal expansion $\alpha$.

The sign change of thermal expansion, $\alpha \propto \partial
S/\partial p$, reveals the locations of extrema of entropy, $S$,
in the phase diagram that originate from the proliferation of
low-energy fluctuations close to a quantum critical point
\cite{Garst05}. In the present context, the close vicinity of two
quantum critical points $H_{c1/c2}$ gives rise to a rich
structure in $\alpha$, see Fig.~\ref{tad}, whose sign changes we
summarized in Fig.~\ref{zeros}. At elevated temperatures,
thermodynamics is not able to resolve the distance between the
critical points, viz. the bandwidth $J_\parallel$ of triplet
excitations. For temperatures $J_\parallel < k_B T \ll J_\perp$,
this is reflected in a broad single peak in the entropy as a
function of magnetic field $S(H)$, or, equivalently, as a
function of the chemical potential $\mu$, indicating the
singlet-triplet level crossing. This peak implies a single sign
change of $\alpha \propto \partial S/\partial J_\perp \propto
\partial S/\partial \mu$ in agreement with the prediction of the
Schottky formula (\ref{schottky}). At $k_B T \simeq J_\parallel$,
however,  thermodynamics starts to resolve the triplet bandwidth
which is reflected in a splitting of the entropy peak into two
separate maxima whose positions approach the critical points
$H_{c1/c2}$ for $T\rightarrow 0$~K. The bifurcation of the
entropy peak is the origin of the rich structure of $\alpha$. The
two maxima and the enclosed minimum of $S(H)$ result in three
consecutive sign changes of $\alpha(H)$ for $k_B T \lesssim
J_\parallel$ as shown in Fig.~\ref{zeros}.

Close to criticality, $H \approx H_{c1/c2}$ and $k_B T \ll
J_\parallel$, the interactions among Jordan-Wigner fermions can
be neglected and the critical model reduces to free
non-relativistic fermions. The resulting singular part of the
free energy has the scaling form
\begin{equation}
F_{cr,n} = |g \mu_B (H-H_{cn})|^{3/2} \phi_n\left(\frac{k_B T}{g
\mu_B (H-H_{cn})}\right)\, , \label{criticalF}
\end{equation}
where $n=1,2$ label the two distinct critical points and the
scaling function $\phi$ was specified in \cite{Sachdev}. Assuming
that the critical fields $H_{c1/c2}$ are smooth functions of
pressure, it follows from (\ref{criticalF}) that $\varepsilon(H)
\propto \partial F_{cr,n}/\partial p \propto
\partial F_{cr,n}/\partial H_{cn}$ approaches the same
characteristic square root behavior as the magnetization in the
zero temperature limit, $\varepsilon(H) \sim \sqrt{|H-H_{cn}|}$;
this is already apparent in the low-$T$ data in Fig.~\ref{MS}.
For the critical thermal expansion the previously announced
divergence $\alpha \sim 1/\sqrt{T}$ is found for $|H-H_{cn}| \ll
T$. Note that a divergent thermal expansion at a quantum critical
point is not prohibited by any fundamental principle. However,
(\ref{criticalF}) also predicts a divergent critical contribution
to the compressibility that might result in a preemptive
first-order transition of the elastic system
\cite{FirstOrderTransition}.

\begin{figure}
\includegraphics[width= .9\linewidth]{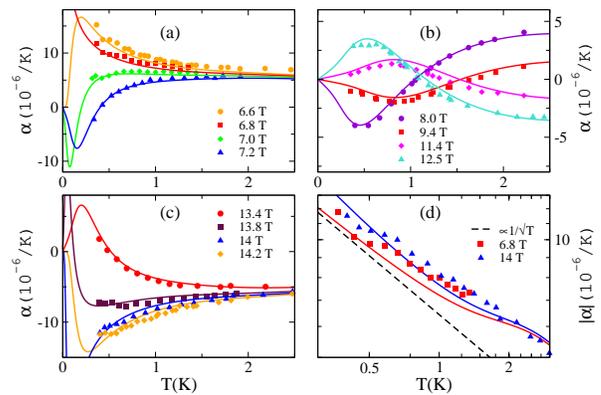}
\caption{\label{exptheo} Panels (a--c) compare representative
experimental data (symbols) of $\alpha(T)$ to theoretical curves
(lines) obtained within the free fermion model (\ref{alphaFF}).
All these curves are fully determined by $H_{c1/c2}$ and a factor
$\gamma$ obtained from a fit to $\varepsilon(H)$ data at
$T=335$~mK (see Fig.~\ref{MS}). Panel (d) shows the thermal
expansion near $H_{c1/c2}$ on a double-logarithmic scale together
with the limiting $1/\sqrt{T}$ behavior (dashed line).}
\end{figure}

In order to capture the complex behavior of $\alpha(T,H)$ for
$k_B T < J_\parallel$ we employ a simple fitting formula
\begin {equation}
\alpha_{FF} = -\gamma \int^\pi_{-\pi} \frac{dk}{2\pi}
\frac{(\mu-\epsilon_k)}{4 k_B T^2\cosh^2((\mu-\epsilon_k)/(2 k_B
T))}\, , \label{alphaFF}
\end{equation}
where $\mu = g \mu_B (2H - H_{c1}-H_{c2})/2$, the dispersion
$\epsilon_k = \frac{1}{2} g \mu_B (H_{c2} - H_{c1}) \cos(k)$, and
$\gamma$ measures the pressure dependence of the singlet-triplet
crossing field, $\gamma = g \mu_B (\partial(H_{c1} +
H_{c2})/\partial p)/(2 V_D)$. Formula (\ref{alphaFF}) follows
from a free fermion model with a pressure dependent chemical
potential $\mu$ and dispersion $\epsilon_k$; note that it
reproduces the qualitatively correct $1/\sqrt{T}$ behavior close
to the critical fields. The fit of $\varepsilon(H)$ at 335~mK,
see Fig.~\ref{MS}, identifies $\gamma \simeq 12 \times 10^{-5}$
with the saturation value of $\varepsilon(H)$ at large fields, in
good agreement with the value obtained from the Schottky fit of
$\alpha(T)$. Having $\gamma$ fixed, Eq.~(\ref{alphaFF})
strikingly reproduces the complex behavior of $\alpha(T)$ in the
low-$T$ limit for all magnetic fields without further adjustment
of any parameter, see Fig.~\ref{exptheo}. In particular,
Eq.~(\ref{alphaFF}) confirms the three consecutive sign changes
of $\alpha$ at lowest $T$ as discussed above. The fit also
demonstrates in Fig.~\ref{exptheo}(a) that in the measured
temperature range the off-critical curve at $H=6.6$~T is even
larger than the critical one at $H_{c1}=6.8$~T, but is expected
to cross the latter at lower $T$. In Fig.~\ref{exptheo}(d) we
show the $\alpha(T)$ curves that are closest to the two critical
fields on a double-logarithmic scale. Whereas $\alpha$ at
$H=6.8$~T is critical, the one at $H=14$~T is slightly
off-critical. The dashed line indicates the limiting $1/\sqrt{T}$
behavior. The deviation of the latter from the critical $H=6.8$~T
fitting curve (solid line) is attributed to certain corrections
to scaling that are still sizeable in the considered temperature
range; note, however, that the simplistic formula (\ref{alphaFF})
does not correctly capture these corrections. The small off-set
between the experimental data and the fit in
Fig.~\ref{exptheo}(d) amounts to a systematic error in the
effective prefactor $\gamma$ at criticality of $\simeq 10\%$
\cite{Anfuso}.

\begin{figure}
\includegraphics[width= .9\linewidth]{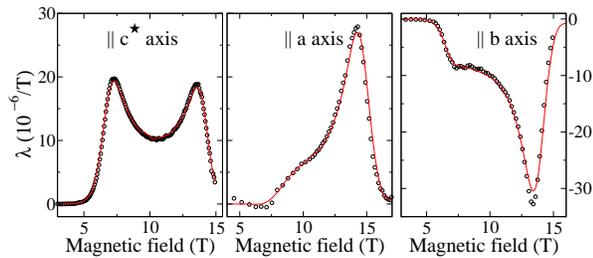}
\caption{\label{msaniso} Anisotropic magnetostriction coefficients
$\lambda$ (symbols) measured along the $c^\star$, $a$, and $b$
axis. The lines are Quantum Monte Carlo fits using the
experimental values of $J_\perp$ and $J_\|$ and two additional
parameters for each direction, which measure the uniaxial
pressure dependencies of $J_\perp$ and $J_\|$~\cite{Anfuso}.}
\end{figure}

Fig.~\ref{msaniso} compares the uniaxial magnetostriction
coefficient $\lambda$ along the $c^\star$ axis with the ones
along the orthogonal $a$ and $b$ directions. While the former is
nearly symmetric with respect to $(H_{c1}+H_{c2})/2$ this
symmetry is missing in the other two directions. This difference
is due to the additional contribution arising from the pressure
dependence of the leg coupling $J_\parallel$. It turns out that
the dependence of $J_\parallel$ on uniaxial pressure
$p_{c^\star}\|c^\star$ is very weak, $|\partial J_\parallel /
\partial p_{c^\star}| \ll |\partial J_\perp /
\partial p_{c^\star}|$, such that its influence can be neglected. The
single dominant $p_{c^\star}$-dependent energy scale $J_\perp$
causes peaks in $\lambda$ at $H_{c1/c2}$ that are of comparable
sizes as $\partial H_{c1}/\partial p_{c^\star} \approx \partial
H_{c2}/\partial p_{c^\star} \propto \partial J_\perp/\partial
p_{c^\star}$ \cite{HcFormulae}. This is not the case for the
orthogonal directions where the respective uniaxial pressure
dependencies of $J_\perp$ and $J_\parallel$ are of similar
magnitudes. In particular, the small relative size of the peaks at
$H_{c1} \approx (J_\perp - J_\parallel)/g \mu_B$ can be
understood as a partial cancellation of the two pressure
dependencies $\partial J_\parallel / \partial p_n \sim \partial
J_\perp / \partial p_n$ yielding a small $\partial
H_{c1}/\partial p_n$, with $n=a,b$. Taking the pressure
dependencies of both, $J_\perp$ and $J_\parallel$, into account,
we find excellent quantitative agreement with Quantum Monte Carlo
simulations of the Hamiltonian (\ref{modelhamiltonian}), solid
lines in Fig.~\ref{msaniso}, whose details will be presented
elsewhere \cite{Anfuso}.

In summary, we measured the magnetostriction and thermal
expansion of \hp . For both physical quantities we find excellent
agreement with calculations based on a two-leg spin ladder
Hamiltonian (\ref{modelhamiltonian}). The thermal expansion
$\alpha$ is critically enhanced as $1/\sqrt{T}$ close to the two
quantum critical points at $H_{c1/c2}$ and shows various sign
changes as a function of $H$ and $T$ signaling entropy extrema in
the phase diagram. This complex behavior of $\alpha$ is
semi-quantitatively explained within a model of free fermions,
and we find quantitative agreement with Quantum Monte Carlo
calculations.

We acknowledge discussions with J.A.\ Mydosh, B.~Thielemann, and
K.~Kiefer. This work was supported by the DFG through SFB~608.
Computer time was allocated through Swedish Grant No.~SNIC
005/06-8.


\begin{thebibliography}{10}
\bibitem{HcFormulae}
The coupling constants can be determined perturbatively from the
critical fields. Up to second order in $J_\parallel$ they are
given by \cite{Reigrotzki94}: $g \mu_B H_{c1} = J_\perp -
J_\parallel + J^2_\parallel/(2 J_\perp)$ and $g \mu_B H_{c2} =
J_\perp + 2 J_\parallel$ with g=2.15 for $H||c^\star$
\cite{patyal90a}.
\bibitem{HcFormulae2} As the Schottky formula (\ref{schottky}) is only valid up to
first order in $J_\parallel$, we neglect here for consistency the
second order in $J_\parallel$ contribution in the formula for
$H_{c1}$ \cite{HcFormulae} yielding $J_\perp/k_B = 13.2$~K and
$J_\parallel/k_B = 3.4$~K.
\bibitem{FirstOrderTransition}
A first order transition is expected when the critical
contribution to compressibility is comparable to the one from the
lattice $\kappa_L$.
An order of magnitude estimate shows that 
this occurs at a temperature $T_c \sim (2\pi)^2 \frac{\gamma^4
V_D^2}{k_B J_\parallel \kappa_L^2} \simeq 5 \times10^{-4}$ K,
where we used $\kappa^{-1}_L \sim 10$ GPa. However, the first
order instability might be prevented by the onset of N\'eel order
due to inter-ladder coupling \cite{ruegg,Anfuso}.
\end{thebibliography}
\end{document}